\documentclass[%
 aip,
 amsmath,amssymb,
preprint,%
]{revtex4-1}

\usepackage{graphicx}
\usepackage{dcolumn}
\usepackage{bm}

\usepackage[utf8]{inputenc}
\usepackage{mathptmx}

\makeatletter
\def\@email#1#2{%
 \endgroup
 \patchcmd{\titleblock@produce}
  {\frontmatter@RRAPformat}
  {\frontmatter@RRAPformat{\produce@RRAP{*#1\href{mailto:#2}{#2}}}\frontmatter@RRAPformat}
  {}{}
}%
\makeatother

\begin{document}


\title{Amplitude Enhancements through rewiring of a non-autonomous delay system}

\author{Kenta Ohira}%
 \affiliation{Graduate School of Informatics, Nagoya University, Nagoya, Japan
}
\author{Toru Ohira}
\affiliation{Graduate School of Mathematics, Nagoya University, Nagoya, Japan}
 
\author{Hideki Ohira}
\affiliation{Graduate School of Informatics, Nagoya University, Nagoya, Japan
}

\date{\today}

\begin{abstract}
Complex systems, such as biological networks, often exhibit intricate rhythmic behaviors that emerge from simple, small-amplitude dynamics in individual components. This study explores how significant oscillatory signals can arise from a minimal system consisting of just two interacting units, each governed by a simple non-autonomous delay differential equation with a recently obtained exact analytical solution. Contrary to the common assumption that large-scale oscillations require numerous units, our model demonstrates that rewiring two units from self-feedback to cross-feedback can generate robust, finite-amplitude oscillations. With time delay, these interacting units produce strongly amplified oscillatory packets compared to self-feedback configurations. Our findings highlight the potential of this minimalistic mechanism for generating complex rhythmic outputs, with implications for oscillatory signal processing and various other applications.
\end{abstract}

\maketitle


\begin{quotation}
Complex systems, such as biological networks, often exhibit intricate rhythmic behaviors (e.g. \cite{foster2004,buzsaki2011,onkar2011,glassmackey1988}). The emergence of these behaviors from a collection of small components, each exhibiting simple, low-amplitude dynamics, is a fascinating phenomenon. In particular, we investigate how large oscillatory signals can arise from a system where individual units generate only small-amplitude, simple dynamics.

A natural assumption is that such amplified emergent behavior requires a large number of interacting units. For example, the sinoatrial node, the primary pacemaker of the heart, typically consists of several thousand to tens of thousands of cells in mammals\cite{Dobrzynski2007,Monfredi2010}. However, we demonstrate that significant oscillatory amplification can emerge with only two units, achieved simply by rewiring from self-feedback to cross-feedback with delays. Systems with feedback delays have been studied in various fields from mathematics, biology, physics, engineering and so on\cite{heiden1979,bellman1963,cabrera1,hayes1950,insperger,kcuhler,longtinmilton1989a,mackeyglass1977,miltonetal2009b,ohirayamane2000,smith2010,stepan1989,stepaninsperger}. While it is well known that systems with delayed feedback or coupled differential equations can produce oscillations, and in some situations leading to divergence (e.g. \cite{Hale1993,Ezzinbi2006,Eremin2021}), our model generates strongly amplified yet finite oscillations.

We begin by introducing a delayed self-feedback unit described by a simple non-autonomous delay differential equation (DDE). Non-autonomous delay dynamical systems, in general, give rise to intriguing phenomena and have gained increasing attention. Particularly with delay, this class of equations is generally considered difficult to solve and has primarily been studied through stability analysis, approximations, or numerical methods \cite{Busenberg1984,Volz1986,Ming1990,Ford2002,Liu2006,Gyori2017,Lucas2018,Kuptsov2020,Herrera2022}. 
However, we present a simple non-autonomous DDE whose exact analytical solution has recently been derived by one of the authors \cite{kentaohira2024}. To the best of our knowledge, this is the first instance where an explicit, analytically exact solution for a non-autonomous DDE has been documented.

We then numerically demonstrate that within a specific parameter range, two such units can significantly enhance oscillation amplitude simply by rewiring from self- to cross-feedback, without altering parameter values.
\end{quotation}

\section{Non-autonomous delayed self-feedback unit}

The fundamental delayed self-feedback unit investigated in this study is described by the following non-autonomous delay differential equation:
\begin{equation}
\frac{dX(t)}{dt} + a t X(t) = b X(t-\tau)
\label{dr}
\end{equation}
where $a > 0$, $b$, and $\tau \geq 0$ are finite real parameters, with $\tau$ representing the time delay. This equation is a special case of a more general linear non-autonomous delay differential equation:
\begin{equation}
\frac{dX(t)}{dt} + f(t) X(t) = g(t) X(t-\tau)
\label{dr_general}
\end{equation}
where $f(t)$ and $g(t)$ are functions of time. Stability analysis, as well as numerical and approximate investigations, have been conducted on this class of equations or systems of such equations \cite{Busenberg1984,Volz1986,Ming1990,Ford2002,Liu2006,Gyori2017,Lucas2018,Kuptsov2020,Herrera2022}. 
However, analyzing these equations, particularly their transient dynamical behaviors, is generally considered challenging.

In previous studies, we analyzed equation (\ref{dr}) and obtained semi-analytical solutions while also observing peculiar frequency resonance phenomena \cite{kentaohira2022,kentaohira2023,kentaohira2025}. 
Furthermore, one of the authors has recently derived the following exact analytical solution for all $t \in \mathbb{R}$ \cite{kentaohira2024}.


\begin{equation}
    X(t) = {\it{C}} \sum_{n = 0}^{\infty} \frac{1}{n!} \left( \frac{b}{a \tau} \right)^n e^{-{1 \over 2} a (t - n \tau)^2}
\label{dsolution}
\end{equation}
where \( {\it{C}} \) is an arbitrary constant. The detailed derivation is provided elsewhere \cite{kentaohira2024}. However, direct substitution readily verifies that this solution satisfies the non-autonomous DDE given in (\ref{dr}). To the best of our knowledge, this is the first instance in which such an explicit, exact solution has been documented for non-autonomous DDEs, and it remains a rare case even among autonomous DDEs.

\begin{figure}[h]
\begin{center}
\includegraphics[height=12cm]{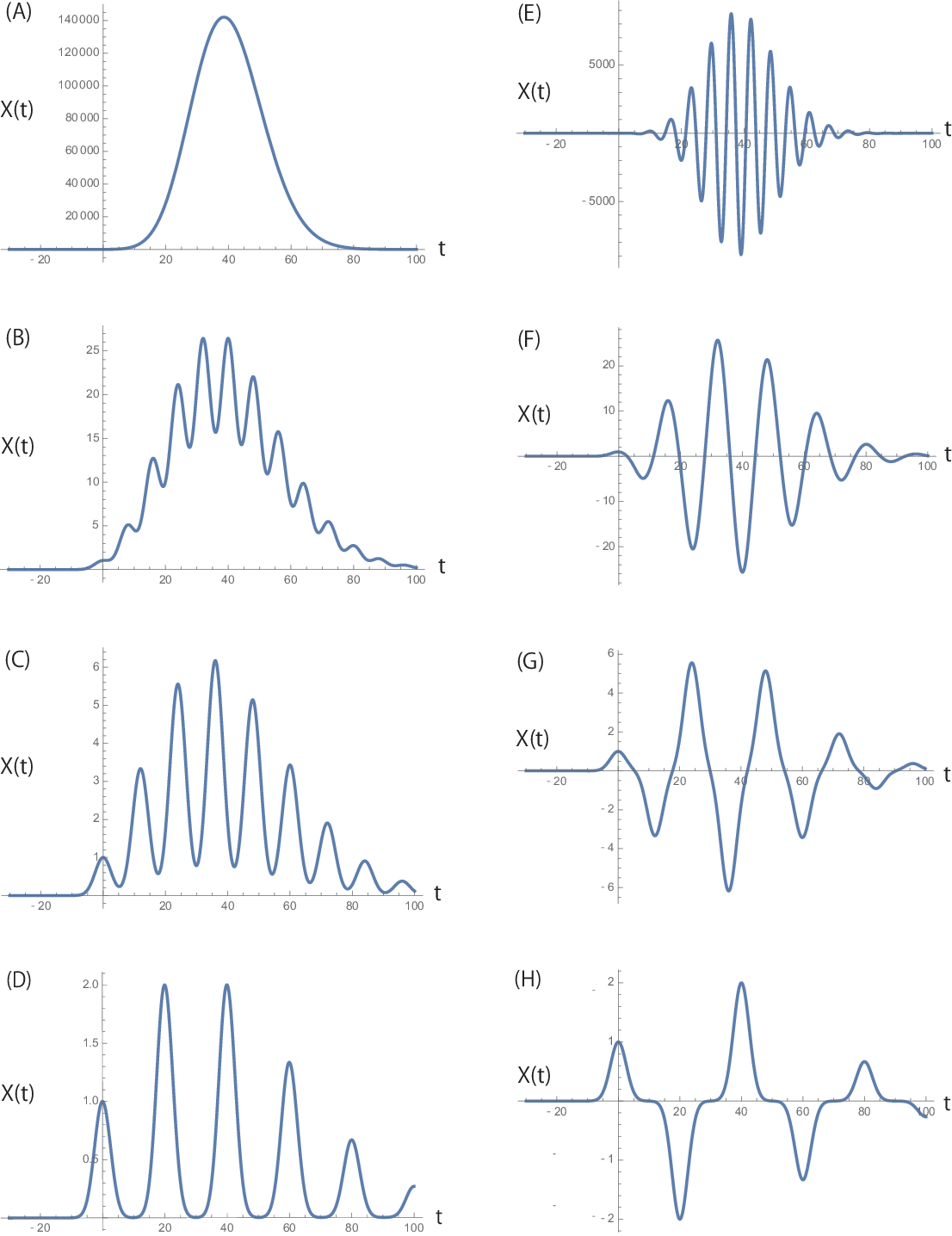}
\caption{Representative dynamics of equation (\ref{dr}) given by the solution (\ref{dsolution}). We set  ${\it{C}} =1$, $a = 0.15$, and sum the series up to $n=500$. Other parameters $(b, \tau)$ are set as follows: (A) (6, 3), (B) (6, 8), (C) (6,12), (D) (6, 20), (E) (-6, 3), (F) (- 6, 8), (G) (- 6, 12), (H) (- 6, 20). }
\label{exactsolution}
\end{center}
\end{figure}

This solution represents a superposition of Gaussian functions centered at \( t = n \tau \), 
with amplitudes given by \( \frac{{\it{C}}}{n!} \left( \frac{b}{a \tau} \right)^n \).
Representative example plots are shown in Figure \ref{exactsolution}.Notably, this solution reveals that increasing the delay suppresses the amplitude of the dynamics.

\section{Two Non-autonomous delayed cross-feedback units}

Next, we consider a system of two non-autonomous units, each governed by equation (\ref{dr}). 
\begin{eqnarray}
&& {dX(t)\over dt} + a t X(t) = b X(t-{\tau_1}) \nonumber \\
&& {dY(t)\over dt} + \alpha t Y(t) = \beta Y(t-{\tau_2})
\label{dr2}
\end{eqnarray}
This system can be interpreted as two independent RC circuits with time-varying resistance and self-feedback through delay lines, as illustrated schematically in Figure 2 (A).

Now, if we rewire the system to introduce cross-feedback instead of self-feedback, the equations become
\begin{eqnarray}
&& {dX(t)\over dt} + a t X(t) = b Y(t-{\tau_1}) \nonumber \\
&& {dY(t)\over dt} + \alpha t Y(t) = \beta X(t-{\tau_2}).
\label{drc}
\end{eqnarray}

By simply rewiring the feedback connections, we retain the same parameter values, as these parameters describe the intrinsic properties of the system components, including the delay lines. The key difference lies in switching the feedback sources, as reflected on the right-hand side of the equations. The corresponding schematic view is shown in Figure 2 (B).
\vspace{1em}

\begin{figure}[h]
\begin{center}
\includegraphics[height=4.5cm]{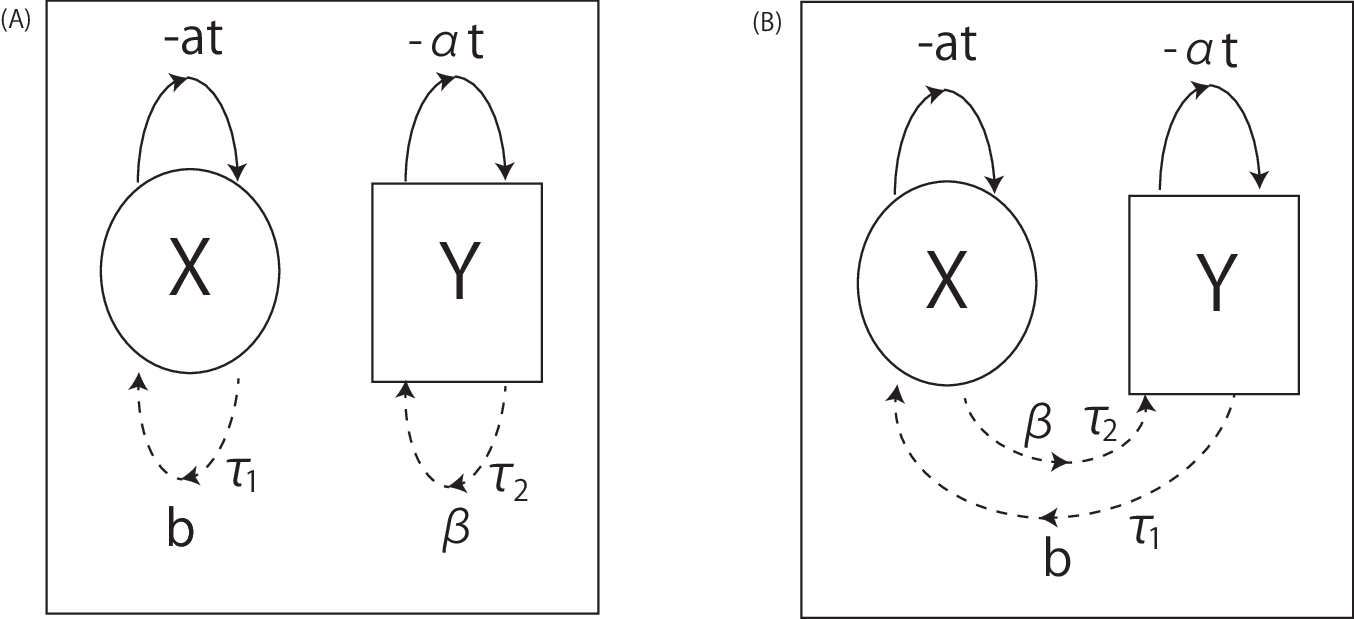}
\caption{Schematic view of (A) self-feedback and (B) cross-feedback systems. Dashed lines indicate delayed transmission lines.}
\label{circuit}
\end{center}
\end{figure}

\section{Stability Analysis}

We begin by discussing the stability analysis of the non-autonomous DDEs (\ref{dr}) and (\ref{drc}).

\subsection{Self-Feedback Case}

The self-feedback equation (\ref{dr}) is a specific case of non-autonomous delay differential equations. As described in the previous section, we have obtained an exact analytical solution (\ref{dsolution}).

Here, we supplement our findings by applying the stability analysis of (\ref{dr}) using the theorems developed by Volz\cite{Volz1986}. By tailoring one of his theorems to be applicable to (\ref{dr}), we arrive at the following corollary:

\vspace{1.5em}

\noindent Corollary A: 
The following non-autonomous DDE,  
\begin{equation}
    \frac{dX(t)}{dt} + a t X(t) = b X(t-\tau)
    \label{drs}
\end{equation}
with \( a \), \( b \), and \( \tau \geq 0 \) as finite real parameters, is asymptotically stable provided that there exist \( t_0 \geq 0 \) and \( \epsilon > 0 \) such that  
\begin{equation}
    e^{\epsilon \tau} |b| \leq (a t - \epsilon).
    \label{drcond}
\end{equation}

\vspace{1.5em}

This stability condition can be satisfied when \( a > 0 \) with finite \( b \) and \( \tau \), by choosing a finite \( \epsilon \), since the right-hand side is a linearly increasing function. A suitable \( t_0 \) can be found to satisfy this inequality.

This result is fully consistent with the exact solution (\ref{dsolution}).

\subsection{Cross-Feedback Case} 

For the cross-feedback case given by (\ref{drc}), we have not yet obtained an exact analytical solution. However, we can still analyze its stability using Volz's theorem on systems of non-autonomous DDEs\cite{Volz1986}. By customizing it for our case, we obtain the following corollary:

\vspace{1.5em}

\noindent Corollary B:  
Consider the system of non-autonomous DDEs:

\begin{equation}
    \frac{dX_i(t)}{dt} + a_i t X_i(t) = \sum_{k=1}^{n} b_{ik} X_k(t-\tau_i),
    \label{drcps}
\end{equation}
where \( i = 1, \dots, n \), and \( a_i, b_{ik}, \tau_i \) are finite real numbers.

This system is asymptotically stable provided that there exist \( t_0 \geq 0 \), \( \epsilon > 0 \), \( \gamma_i > 0 \), and \( T \geq \max[\tau_i] \) (for all \( i \)) such that

\begin{equation}
    e^{\epsilon T} \sum_{k=1}^{n} \gamma_k |b_{ik}| \leq \gamma_i (a_i t - \epsilon),
    \label{drcgcond}
\end{equation}
for all \( t \geq t_0 \) and \( i = 1, \dots, n \).

\vspace{1.5em}

Applying this theorem to our specific case (\ref{drc}), we set  
\( n = 2 \), \( a_1 = a \), \( a_2 = \alpha \), \( b_{11} = 0 \), \( b_{12} = b \), \( b_{21} = \beta \), and \( b_{22} = 0 \).  
The two stability conditions then reduce to the following inequalities:

\begin{eqnarray}
    e^{\epsilon T} \gamma_2 |b| &\leq \gamma_1 (a t - \epsilon), \\
    e^{\epsilon T} \gamma_1 |\beta| &\leq \gamma_2 (\alpha t - \epsilon).
    \label{drccond}
\end{eqnarray}

If we choose \( T \geq \max[\tau_1, \tau_2] \) and set finite values for \( \epsilon > 0 \) and \( \gamma_1 = \gamma_2 = 1 \), we can clearly satisfy these conditions for all \( t \geq t_0 \) when both \( a > 0 \) and \( \alpha > 0 \), with a suitably chosen \( t_0 \).

These theoretical results align well with our numerical simulations of equation (\ref{drc}), which confirm that the system remains asymptotically stable when both \( a > 0 \) and \( \alpha > 0 \).

\section{Amplitude Enhancement Phenomena}

We have discovered a peculiar phenomenon of strong amplitude enhancement that arises when transitioning from the self-feedback system (\ref{dr2}) to the cross-feedback system (\ref{drc}) for certain parameter ranges. We investigate this phenomenon numerically and compare the dynamic behaviors of both systems.

\begin{figure}
    \begin{center}
        \includegraphics[height=19.5cm]{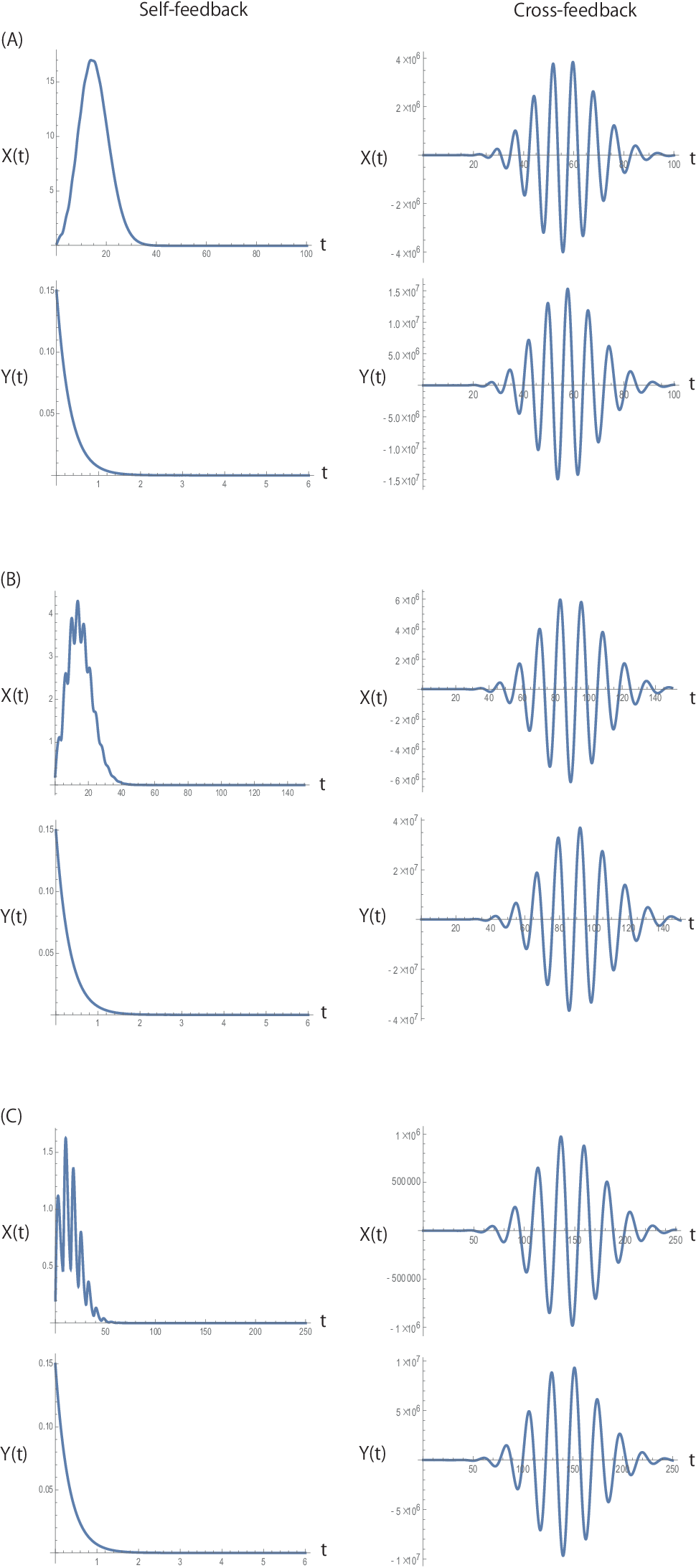}
        \caption{Examples of amplitude enhancement phenomena: The left column shows self-feedback dynamics (\ref{dr2}), while the right column shows cross-feedback dynamics (\ref{drc}). Note the difference in vertical-axis scales. The initial conditions are set at constant values \( X_0 = 0.2 \) and \( Y_0 = 0.15 \) for \( t \in [-\tau,0] \). Parameters are set as \( a = 0.2 \), \( \alpha = 0.001 \), \( b = 3.0 \), \( \beta = 3.0 \), and \( \tau_2 = 0.01 \). The values of \( \tau_1 \) vary: (A) \( \tau_1 = 2.0 \), (B) \( \tau_1 = 3.5 \), (C) \( \tau_1 = 7.5 \).}
        \label{dyc}
    \end{center}
\end{figure}

\begin{figure}
    \begin{center}
        \includegraphics[height=5cm]{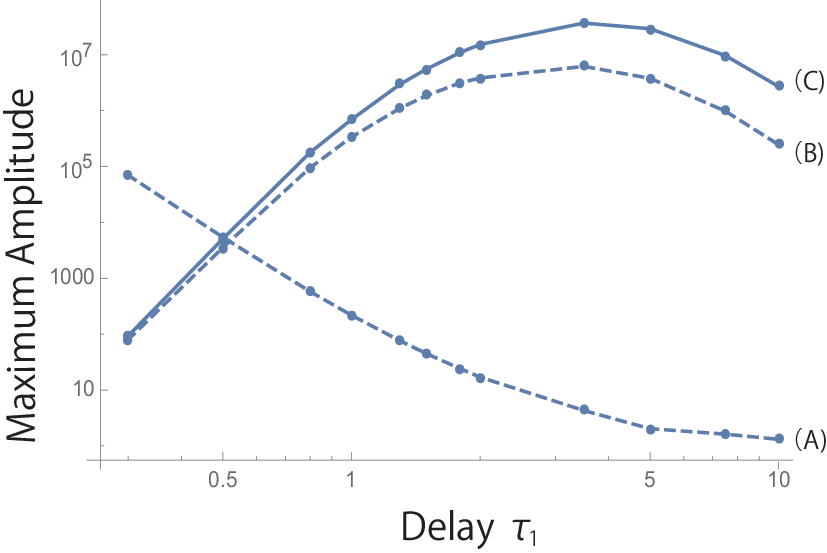}
        \caption{Log-log plots of the maximum amplitude as a function of delay \( \tau_1 \) for self-feedback (\( X(t) \)) (A), \( Y(t) \) (too small to be shown), and cross-feedback (\( X(t) \)) (B), \( Y(t) \) (C). The other parameters remain the same as in Fig 3.}
        \label{maxamp}
    \end{center}
\end{figure}

The initial conditions are set to constant values: \( X_0 = 0.2 \) and \( Y_0 = 0.15 \) for \( t \in [-\tau,0] \). Figure~\ref{dyc} shows representative plots comparing the dynamics of the self-feedback system (\ref{dr2}) and the cross-feedback system (\ref{drc}). For these plots, we fix the parameters as follows: \( a = 0.2 \), \( \alpha = 0.001 \), \( b = 3.0 \), \( \beta = 3.0 \), and \( \tau_2 = 0.01 \). Additionally, we plot the maximum amplitude values as a function of the delay \( \tau_1 \) for both systems in Figure~\ref{maxamp}.

Based on these results, we observe the following:

\begin{enumerate}
    \item[(i)] For the self-feedback system (\ref{dr2}), increasing the delay \( \tau_1 \) suppresses the amplitude of \( X(t) \). With sufficiently large \( \tau_1 \), transient oscillatory behaviors appear. Meanwhile, \( Y(t) \) continues to decay monotonically since \( \tau_2 \) remains fixed.
    
    \item[(ii)] In contrast, for the cross-feedback system (\ref{drc}), increasing the delay induces an increase in the amplitude of both \( X(t) \) and \( Y(t) \).
    
    \item[(iii)] A crossover point exists where the amplitude of \( X(t) \) in the cross-feedback system surpasses that of the self-feedback system. For this parameter setting, it occurs around \( \tau_1 \approx 0.5 \). For \( Y(t) \), the amplitude in the cross-feedback system is consistently larger for all \( \tau_1 > 0 \).
    
    \item[(iv)] When the delay exceeds this crossover value, the amplitude in the cross-feedback system becomes significantly larger compared to the corresponding self-feedback system, purely due to the rewiring.
    
    \item[(v)] As seen in Figures~\ref{dyc} and~\ref{maxamp}, the enhancement in amplitude due to cross-feedback can reach as high as \( 10^8 \).
\end{enumerate}

Thus, when a sufficient delay is present, one can extract significantly enhanced amplitude signals from the cross-feedback system, even when the individual self-feedback units produce only small signals. This enhancement arises solely from the rewiring.

We also note that similar enhancement effects were observed for a range of parameter values, not just the specific ones presented here.

Furthermore, while delays often lead to instability or divergence in dynamic systems, here we obtain signals that are both significantly large and finite. As demonstrated in our stability analysis, these signals decay to zero in the long-time limit. This property makes the system practical for implementation in real circuits and other applications.
\clearpage

\clearpage

\section{Discussion}

We first presented an exact analytical solution for a simple non-autonomous delay differential equation (DDE), recently obtained by one of the authors. Building upon this self-feedback unit, we identified a novel physical phenomenon, amplitude enhancement, achieved through a simple rewiring to a cross-feedback configuration. 

Through numerical investigations of a concrete example, we demonstrated that even with just two interacting units, each producing only a small individual output, the system can generate oscillatory signals with significantly enhanced amplitudes. The dynamics in such cases typically take the form of oscillatory wave packets. Conversely, if the goal is to generate such oscillatory packets, this can be achieved by allowing self-feedback units with small outputs to interact, as demonstrated in this study.

Future investigations should address the following:
\begin{enumerate}
    \item[(i)] Derivation of an exact solution for the non-autonomous cross-feedback DDEs (\ref{drc}).
    \item[(ii)] Experimental realization and confirmation of the amplitude enhancement phenomenon.
\end{enumerate}

Regarding the second point, oscillatory wave packets arise in various scientific and engineering domains and are essential for applications in information processing and signal engineering (e.g., \cite{daubechies,ahmad,fortier,freeman}). We hope that our proposed model and its associated enhancement phenomena will serve as a contribution to these fields.

\begin{acknowledgments}
This work was supported by the Yocho-gaku Project sponsored by Toyota Motor Corporation,  
JSPS Topic-Setting Program to Advance Cutting-Edge Humanities  
and Social Sciences Research (Grant Number JPJS00122674991),  
JSPS KAKENHI (Grant Number 19H01201), and the Research Institute for Mathematical Sciences,  
an International Joint Usage/Research Center located at Kyoto University.
\end{acknowledgments}

\end{document}